\documentclass[aps,prc,twocolumn,superscriptaddress,nofootinbib,preprintnumbers]{revtex4-2}

\usepackage{xcolor}
\usepackage{amsmath}
\usepackage{amssymb}
\usepackage{graphicx}
\usepackage{multirow}
\usepackage{orcidlink} 
\begin{document}
\preprint{}
% Use the \preprint command to place your local institutional report
% number in the upper righthand corner of the title page in preprint mode.
% Multiple \preprint commands are allowed.
% Use the 'preprintnumbers' class option to override journal defaults
% to display numbers if necessary
%\preprint{}

%Title of paper
\title{On the symmetry energy expansion and the peak value of the bulk viscosity}
% repeat the \author \affiliation  etc. as needed
% \email, \thanks, \homepage, \altaffiliation all apply to the current
% author. Explanatory text should go in the []'s, actual e-mail
% address or url should go in the {}'s for \email and \homepage.
% Please use the appropriate macro foreach each type of information

% \affiliation command applies to all authors since the last
% \affiliation command. The \affiliation command should follow the
% other information
% \affiliation can be followed by \email, \homepage, \thanks as well.

\author{Steven P.~Harris}
%\email{stharr@iu.edu}
\affiliation{Center for the Exploration of Energy and Matter and Department of Physics,
Indiana University, Bloomington, IN 47405, USA}
\date{July 20, 2025}
\begin{abstract}
The symmetry energy expansion is a useful way to parametrize the properties of dense matter near nuclear saturation density, and much work has been done to connect physical quantities like the neutron star radius and the core-crust transition density to the symmetry energy parameters.  In this work, I connect the weak-interaction-driven bulk viscosity in neutron-proton-electron ($npe$) matter to the symmetry parameters by calculating the susceptibilities of dense matter in terms of the symmetry energy.  I use this result to calculate the resonant-peak value of the bulk viscosity as a function of density, finding that it strongly depends on $L$, as does the minimum bulk-viscous dissipation timescale.  Also resulting from this calculation is a formula for finding the conformal points of the zero-temperature equation of state.  Finally, I determine for which values of the symmetry parameters the maximum r-mode-stable rotation frequency of an $npe$-matter neutron star is smaller than the Kepler frequency, in the high-temperature conditions where bulk viscosity is the dominant dissipation mechanism.
\end{abstract}
\maketitle
%%%%%%%%%%%%%%%%%%%%%%%%%%%%%%%%%%%%%%%%%%%%%%%%%%%%%%
\section{Introduction}\label{intro}
The equation of state (EoS) of dense matter describes thermodynamic properties of the matter, such as the energy density $\varepsilon$ and pressure $P$, as a function of its baryon density $n_B$ and temperature $T$ \cite{Oertel:2016bki}. The EoS inherently applies only to matter that is in thermal equilibrium, but the assumption of thermal equilibrium is a good one for matter in neutron stars, neutron star mergers, and supernovae  (with the exception of, in some conditions, neutrinos \cite{Foucart:2022bth,Espino:2023dei,Endrizzi:2019trv} and perhaps feebly-interacting particles representing physics beyond the standard model \cite{Caputo:2024oqc,Dev:2023hax,Diamond:2023cto}).  Neutron star matter is multicomponent, containing (at the very least) neutrons, protons, and electrons, and while weak interactions attempt to drive the matter towards beta (chemical) equilibrium, dense matter is likely to be out of beta equilibrium in a wide number of situations of current interest \cite{Espino:2023dei,Hammond:2021vtv,Arras:2018fxj,Fernandez:2005cg}.

Great effort has been undertaken to calculate and measure the EoS of dense matter, as it directly controls the allowed masses and radii of neutron stars \cite{MUSES:2023hyz,Oertel:2016bki,Rutherford:2024srk,Annala:2021gom,Legred:2021hdx}, the extent to which they are tidally deformed in the inspiral phase of a neutron star merger \cite{MUSES:2023hyz,Annala:2021gom,Chatziioannou:2020pqz}, and the oscillation spectrum of both isolated neutron stars \cite{Kunjipurayil:2022zah,Roy:2023gzi} and neutron star merger remnants \cite{Rezzolla:2016nxn,Takami:2014zpa,Bauswein:2011tp}.  A natural starting point for calculating the EoS is nuclear saturation density, $n_0\equiv 0.16 \text{ fm}^{-3}$, as this density is reached in atomic nuclei, providing a wealth of experimental input \cite{MUSES:2023hyz,Lattimer:2023rpe}.  In this approach, one Taylor expands the EoS around the conditions encountered in nuclei [nuclear saturation density and symmetric nuclear matter ($x_p=1/2$)] and introduces the symmetry energy: the energy difference between symmetric and pure neutron matter.  The coefficients of the Taylor expansion have become the object of much research and the lower order terms are fairly well constrained \cite{Lattimer:2023rpe,Li:2019xxz}.    

While the EoS describes equilibrium properties of dense matter, astrophysical environments usually are dynamic, with energy, momentum, and heat flow from one part of the system to another.  Transport properties describe how these flows proceed and, at their heart, involve the rate of some process, whether it is energy/momentum exchange in a scattering process or flavor conversion in a weak interaction \cite{Schmitt:2017efp}.  Bulk viscosity has long been a key transport property in neutron star contexts, from its ability to damp radial oscillations of newly born neutron stars \cite{1966ApJ...145..514M,Gusakov:2005dz}, to its importance in stabilizing r-mode oscillations of rotating neutron stars \cite{Andersson:1998ze,Alford:2010fd,Kolomeitsev:2014gfa,Alford:2011pi} and its role in damping other neutron star nonradial oscillation modes \cite{Zhao:2025pgx,Andersson:2019mxp,2012MNRAS.422.3327P,1990ApJ...363..603C}, culminating in its possible impact on neutron star mergers \cite{Alford:2017rxf,Alford:2019qtm,Alford:2020lla,Most:2021zvc,Alford:2021lpp,Alford:2022ufz,Hernandez:2024rxi,Alford:2020pld,Ghosh:2025wfx,Arras:2018fxj,Harris:2024ssp,Alford:2023gxq,Celora:2022nbp} and recent implementation in numerical simulations of these mergers \cite{Chabanov:2023blf,Chabanov:2023abq,Most:2022yhe,Espino:2023dei}.  A history and a pedagogical discussion of bulk viscosity is presented in \cite{Harris:2024evy}.

However, in neutron star contexts there is often little linkage made between the EoS and transport: typically the two are calculated in entirely different frameworks.  This often seems natural, because the EoS is dominated by strong interaction physics, while transport may be connected to a weak or electromagnetic interaction.  Weak interactions, with their small cross sections, contribute negligibly to the EoS and electromagnetism contributes only at the lowest densities encountered in neutron stars.  Some relevant rates for transport do involve strong interactions, but often the strong interaction is modeled with one-pion exchange \cite{Friman:1979ecl} and in the mean field approximation (a common technique used in EoS construction) the pion contribution to the EoS vanishes \cite{Glendenning:1997wn}.  Connecting the EoS and approaches used in transport calculations in neutron stars is not necessarily easy.  

Some connections between transport and the EoS have been made recently.  Efforts have been made to relate the direct Urca threshold, a key aspect of the Urca rate, to the symmetry energy \cite{Lattimer:1991ib,Cavagnoli:2011ft,Steiner:2006bx,Lopes:2014wda,Sarkar:2023xjd}.  Additionally, weak interaction rates in dense matter pick up strong-interaction corrections.  For example, the modified Urca process, where a neutron undergoing weak decay interacts via the strong interaction with a bystander nucleon, represents an important correction to the purely weak direct Urca process \cite{Yakovlev:2000jp,Schmitt:2017efp,Shternin:2018dcn,Alford:2024xfb}.  A recent work calculated neutrino opacities with many-body corrections and ensured that the strong interaction physics that went into the opacities was consistent with the EoS used \cite{Lin:2022lug}.   In \cite{Vidana:2012ex,Tu:2025vew}, the bulk and shear viscosity coefficients were fit to functions of the symmetry energy slope parameter, albeit incompletely, with an eye towards understanding the relationship between the symmetry energy and the r-mode instability window.  

Finally, two recent works by Yang \textit{et al.}~have explicitly focused on the bulk viscosity \cite{Yang:2023ogo,Yang:2025yoo}.  In \cite{Yang:2023ogo}, relativistic mean field (RMF) theory EoSs were constructed with different symmetry energy parameters and those parameters were found to correlate with the zero-frequency bulk viscosity.  In \cite{Yang:2025yoo}, Yang \textit{et al.}~constructed expressions for the zero-frequency bulk viscosity and the beta-equilibration rate in terms of the symmetry energy more generally.

In this work, I connect the resonant-peak value of the bulk viscosity of neutron-proton-electron ($npe$) matter to the symmetry parameters and explore the impact of current constraints on those coefficients on transport in dense matter.  In Sec.~\ref{sec:sym_energy}, I discuss the symmetry energy expansion and constraints on its coefficients.  The susceptibilities of $npe$ matter are derived in Sec.~\ref{sec:bulk_viscosity}, and in Sec.~\ref{sec:implications}, I derive the resonant-peak value of the bulk viscosity and use that result to calculate the minimum bulk-viscous dissipation timescale for small-amplitude density oscillations.  Finally, I apply the peak bulk viscosity result to the r-mode instability window to calculate the maximum neutron star rotation frequency that is stable to r-modes.
 
I work in natural units, where $\hbar=c=k_B=1$.  All data presented in the figures can be found in the Zenodo repository \cite{harris_2025_15557998}.
%%%%%%%%%%%%%%%%%%%%%%%%%%%%%%%%%%%%%%%%%%%%%%%%%%%%%%%%
\section{The symmetry energy expansion}\label{sec:sym_energy}
Near nuclear saturation density, matter is expected to be $npe$ matter: a strongly interacting fluid of neutrons and protons with a free Fermi gas of electrons.  In this article, I will focus only on the zero temperature EoS.  At a given baryon density $n_B\equiv n_n+n_p$ the proton fraction $x_p\equiv n_p/n_B$ assumes a specific value $x_p^{\text{eq.}}$ in beta equilibrium.  However, in many astrophysical situations of relevance, the system can be out of beta equilibrium by an amount 
\begin{equation}
    \delta\mu\equiv \mu_n-\mu_p-\mu_e.
\end{equation}
In zero-temperature $npe$ matter, the first law of thermodynamics is \cite{Harris:2024evy}
\begin{equation}
    \mathop{d\left(\dfrac{\varepsilon}{n_B}\right)}=\dfrac{P}{n_B^2}\mathop{dn_B}-\delta\mu\mathop{dx_p}.  
\end{equation}
Then, the pressure and $\delta\mu$ can be obtained from the derivatives
\begin{equation}
    P = n_B^2\dfrac{\partial\left(\varepsilon/n_B\right)}{\partial n_B}\bigg\vert_{x_p},\quad
    \delta\mu = -\dfrac{\left(\varepsilon/n_B\right)}{\partial x_p}\bigg\vert_{n_B}.\label{eq:P_dmu_thermo}
\end{equation}

As the lepton-baryon interaction is negligible, the energy per baryon can be split into baryonic and leptonic components
\begin{equation}
        \dfrac{\varepsilon}{n_B} = \dfrac{E_{\text{nuc}}}{N_B}+\dfrac{\varepsilon_e}{n_B}.\label{eq:original_EN}
\end{equation}
The leptons (electrons) are treated as a free Fermi gas and their mass is taken to be zero.  The treatment of the baryonic sector is explained below.

The idea of the symmetry energy expansion is to Taylor expand the energy per baryon $E_{\text{nuc}}/N_B$ around a point that is understood very well, namely the conditions encountered in nuclei: $\left\{n_B,x_p\right\} = \left\{n_0,1/2\right\}$.  For historical reasons, the expansion is in powers of $(n_B-n_0)/(3n_0)$ and $(1-2x_p)$.  Neutron star matter is not close to symmetric, so carrying forward with the expansion around $x_p=1/2$ is not appropriate.  It is conventional to terminate the expansion in $(1-2x_p)$ at quadratic order and to assume that this $x_p$-dependence encapsulates the entirety of the $x_p$-dependence of the baryonic part of the EoS.  This is called the \textit{parabolic approximation} and there is good evidence to support its approximate validity \cite{Yang:2025yoo,Wellenhofer:2016lnl,Drischler:2013iza}.  Thus, the energy per baryon in the nuclear sector is 
    \begin{equation}
        \dfrac{E_{\text{nuc}}}{N_B} = \dfrac{E_{\text{nuc}}}{N_B}\left(n_B,x_p=\dfrac{1}{2}\right)+S(n_B)\left(1-2x_p\right)^2.
    \end{equation}
The coefficient $S(n_B)$ of the quadratic term is called the symmetry energy.  Now, both the symmetry energy and the symmetric matter expressions are Taylor expanded in density around $n_0$.  The symmetry energy is given by
\begin{align}
    S(n_B) &= J + \dfrac{L}{3}\left(\dfrac{n_B-n_0}{n_0}\right)+\dfrac{K_{\text{sym}}}{18}\left(\dfrac{n_B-n_0}{n_0}\right)^2\nonumber\\
    &+\dfrac{Q_{\text{sym}}}{162}\left(\dfrac{n_B-n_0}{n_0}\right)^3\label{eq:S(n)}
\end{align}
and the energy per baryon of symmetric nuclear matter is
\begin{align}
    \dfrac{E_{\text{nuc}}}{N_B}\left(n_B,x_p=\dfrac{1}{2}\right) &= m-B_{\text{sat}}+\dfrac{K}{18}\left(\dfrac{n_B-n_0}{n_0}\right)^2\nonumber\\
    &+\dfrac{Q}{162}\left(\dfrac{n_B-n_0}{n_0}\right)^3.
\end{align}
%%%%%%%%%%%%%%%%%%%%%%%%%%%%%%%%
\subsection{Constraints on the Symmetry Parameters}
The coefficients $J$, $L$, $m-B_{\text{sat}}$, and $K$, and to a much lesser extent $K_{\text{sym}}$, $Q$, and $Q_{\text{sym}}$, are well studied, and many constraints have been placed on them.  I will briefly review the constraints, but more extensive studies should be consulted for full details \cite{MUSES:2023hyz,Lattimer:2023rpe,Burgio:2024xpb,Li:2019xxz,Oertel:2016bki,Newton:2011dw}.

The quantity $m-B_{\text{sat}}$ is the nucleon mass (in vacuum) minus the binding energy in symmetric nuclear matter at $n_0$.  With reference to the semi-empirical mass formula for nuclei, the binding energy is determined to be $B_{\text{sat}} = 16\pm 1$ MeV \cite{Lattimer:2023rpe,MUSES:2023hyz} (though see the critique of Atkinson \textit{et al.}~\cite{Atkinson:2020yyo}).  The incompressibility $K$ is measured from giant monopole resonances in heavy nuclei, yielding $K \approx 230\pm 20$ MeV \cite{Lattimer:2023rpe}.  The skewness $Q$ is essentially unconstrained: Burgio \textit{et al.}~\cite{Burgio:2024xpb} mention an estimate $-500\text{ MeV}<Q<300\text{ MeV}$, while Li \textit{et al.}~\cite{Li:2019xxz} quote $-800\text{ MeV}<Q<400\text{ MeV}$.

The symmetry energy coefficients are simultaneously constrained in the laboratory by nuclear mass data, neutron skin measurements, and measurements of collective flow in relativistic heavy ion collisions \cite{Lattimer:2023rpe}.  Relatively strong constraints are possible on the lower order coefficients, but the higher order coefficients are subject to vast uncertainties.  The symmetry energy at saturation density, $J$, is well constrained, lying in the range $29\text{ MeV} < J < 34 \text{ MeV}$ \cite{MUSES:2023hyz}  The slope parameter $L$ is much less certain.  A collection of nuclear experiments have constrained $43\text{ MeV} < L < 75 \text{ MeV}$ \cite{MUSES:2023hyz}. 
 Drischler \textit{et al.}~calculated $L$ with chiral effective field theory ($\chi$EFT) and found $L=59.8\pm 4.1$ MeV \cite{Drischler:2020hwi}.  The results of the PREX-II experiment \cite{PREX:2021umo} lead to the prediction of a particularly large value $L=106\pm 37$ MeV \cite{Reed:2021nqk}.   However, Essick \textit{et al.}~combined the PREX-II experiment results with $\chi$EFT and astrophysical data and inferred $38\text{ MeV} < L < 67 \text{ MeV}$ (90\% credible interval), which is in mild tension with the large value predicted by PREX-II \cite{Essick:2021kjb}.  A more recent calculation combining neutron star data and $\chi$EFT shows $44\text{ MeV} < L < 70 \text{ MeV}$ (95\% credible interval) \cite{Lim:2023dbk}.  Finally, another very recent calculation reanalyzed the PREX-II and CREX results \cite{PREX:2021umo,CREX:2022kgg} with a different set of energy density functionals and found that PREX-II predicts  $38\text{ MeV} < L < 160 \text{ MeV}$ and CREX predicts  $1\text{ MeV} < L < 92 \text{ MeV}$ (95\% credible interval) \cite{Koehn:2024set}.
 
 As mentioned, $K_{\text{sym}}$ is even less certain.  Essick \textit{et al.}~find $-221\text{ MeV} < K_{\text{sym}} < 27 \text{ MeV}$ \cite{Essick:2021kjb}, while Li \textit{et al.}~mention a possible range $-400\text{ MeV}<K_{\text{sym}}<100\text{ MeV}$ \cite{Li:2019xxz}.  Most analyses lead to a negative $K_{\text{sym}}$, though recently a few models with large $K_{\text{sym}}$ were produced with the goal of reconciling the PREX and CREX measurements \cite{Reed:2023cap}.  Finally, $Q_{\text{sym}}$ is virtually unknown: Li \textit{et al.}~suggest $-200 \text{ MeV}<Q_{\text{sym}}<800\text{ MeV}$ \cite{Li:2019xxz}.

In producing an EoS, for example, one built on a Skyrme model \cite{Dutra:2012mb} or an RMF theory \cite{Dutra:2014qga}, one introduces parameters that are then fit to experiment including properties of finite nuclei and of bulk nuclear matter \cite{Glendenning:1997wn}.  Each of these EoSs has its own predicted set of symmetry coefficients, and correlations between the symmetry coefficients are observed \cite{Lattimer:2023rpe}.  Fitting nuclear masses leads to a strong linear correlation between $J$ and $L$ \cite{Lattimer:2023rpe}.  Several correlations are listed by Li \& Magno \cite{Li:2020ass}, including ones derived from Skyrme and RMF theories \cite{Tews:2016jhi,Mondal:2017hnh} and from properties of Fermi liquid theory \cite{Holt:2018uug}.  In this paper, I will use the relations given by Tews \textit{et al.}~\cite{Tews:2016jhi}
\begin{subequations}
\begin{align}
L &= 11.969J-(319.55\pm 41.56)\text{ MeV}\label{eq:L_of_J}\\
K_{\text{sym}} &= 3.501 L - (305.67\pm 56.59)\text{ MeV}\label{eq:Ksym_of_L}\\
Q_{\text{sym}} &= -6.443L + (708.74\pm 171.34)\text{ MeV}.\label{eq:Qsym_of_L}
\end{align}
\end{subequations}
The linear relationships are based around a median relationship with a plus/minus offset added to encompass 95.4\% of the studied EoSs (the EoSs were first screened by Tews \textit{et al.}~for compatibility with experiment).   Similar linear relationships were seen in a different class of EoSs developed by Oyamatsu \cite{Oyamatsu:2022twl}.
%%%%%%%%%%%%%%%%%%%%%%%%%%%%%%%%%%%%%%%%%%%%%%
\subsection{The EoS in terms of Symmetry Parameters}
From Eq.~\ref{eq:original_EN}, the energy density of dense matter, not necessarily in beta equilibrium, is
\begin{align}
    \varepsilon\left(n_B,x_p\right) &= n_B\left[\dfrac{E_{\text{nuc}}}{N_B}\left(n_B,x_p=\dfrac{1}{2}\right) +S(n_B)\left(1-2x_p\right)^2\right]\nonumber\\
    &+ \dfrac{(3\pi^2x_pn_B)^{4/3}}{4\pi^2}\label{eq:edens}
\end{align}
and the pressure, calculated via Eq.~\ref{eq:P_dmu_thermo}, is
\begin{widetext}
\begin{align}
P\left(n_B,x_p\right) &= \dfrac{n_0}{3}\left(\dfrac{n_B}{n_0}\right)^2\bigg\{   
\dfrac{K}{3}\left(\dfrac{n_B-n_0}{n_0}\right)+\dfrac{Q}{18}\left(\dfrac{n_B-n_0}{n_0}\right)^2\label{eq:Pressure}\\
&+\left[L+\dfrac{K_{\text{sym}}}{3}\left(\dfrac{n_B-n_0}{n_0}\right)+\dfrac{Q_{\text{sym}}}{18}\left(\dfrac{n_B-n_0}{n_0}\right)^2\right]\left(1-2x_p\right)^2\bigg\}+\dfrac{\left(3\pi^2x_pn_B\right)^{4/3}}{12\pi^2}.\nonumber
\end{align}
\end{widetext}
From Eq.~\ref{eq:P_dmu_thermo}, one sees
\begin{equation}
     \delta\mu\left(n_B,x_p\right) = 4S(n_B)(1-2x_p)-\left(3\pi^2n_Bx_p\right)^{1/3}.\label{eq:dmu_specifically}
\end{equation}
By setting\footnote{At finite temperature, if the matter remains neutrino-transparent, this beta-equilibrium condition picks up finite-temperature corrections \cite{Alford:2018lhf,Alford:2021ogv,Alford:2023gxq,Hammond:2021vtv}, but I ignore these here.} $\delta\mu=0$, one obtains an equation for the proton fraction in beta equilibrium $x_p^{\text{eq.}}$
\begin{equation}
    64S(n_B)^3\left(1-2x_p^{\text{eq.}}\right)^3=3\pi^2n_Bx_p^{\text{eq.}}.\label{eq:beta_eq_condition}
\end{equation}
Evidently, the proton fraction is set by the symmetry energy at a particular density (as is well known \cite{Lattimer:1991ib,Page:2004fy,Li:2019xxz,Horowitz:2002mb,Tsang:2023vhh}).  Through Eq.~\ref{eq:beta_eq_condition}, the energy density and pressure (Eqs.~\ref{eq:edens} and \ref{eq:Pressure}) can be evaluated in beta equilibrium.  At saturation density, $\varepsilon = n_0\left\{m-B_{\text{sat}}+J\left[1-x_p^{\text{eq.}}-2(x_p^{\text{eq.}})^2\right]\right\}$ and $P = \left(n_0/3\right)\left(1-2x_p^{\text{eq.}}\right)\left[L+\left(3J-2L\right)x_p^{\text{eq.}}\right]$.  The proton fraction is quite small in beta equilibrated matter at $n_0$, and in the limit $x_p\rightarrow 0$, these expressions agree with the neutron matter expressions (e.g.,~Ch.~7 in \cite{Schaffner-Bielich:2020psc}) and the beta-equilibrium pressure expression in \cite{Lattimer:2012xj}.

The proton fraction in beta equilibrium is a function of density (c.f.~Fig.~2 in \cite{Harris:2024evy}).  An interesting potential feature in this curve is a conformal point: a point where $x_p^{\text{eq.}}(n_B)$ reaches a local maximum (or minimum).  Conformal points have important implications for the bulk viscosity \cite{Harris:2024evy,Harris:2024ssp}.  Eq.~\ref{eq:beta_eq_condition} can be used to find conformal points in the EoS. 
 Differentiating Eq.~\ref{eq:beta_eq_condition} with respect to density, solving for $\mathop{dx_p^{\text{eq.}}}/\mathop{dn_B}$ and then setting it to zero yields the condition
\begin{equation}
    3n_B^{\text{conf.}} \dfrac{\mathop{dS(n_B^{\text{conf.}})}}{\mathop{dn_B^{\text{conf.}}}}=S(n_B^{\text{conf.}}).\label{eq:conformalpointseqn}
\end{equation}
This is a polynomial equation in $n_B^{\text{conf.}}$ which, when solved for $n_B^{\text{conf.}}$, gives the location of all of the conformal points of the EoS for a particular choice of symmetry parameters.  For example, if one terminates the symmetry expansion at linear order in $\alpha\equiv (n_B-n_0)/n_0$, one conformal point
\begin{equation}
    \alpha^{\text{conf.}}=-\dfrac{3}{2}\left(\dfrac{L-J}{L}\right) \quad \text{[linear order in $S(n_B)$]}\label{eq:conformal_linear}
\end{equation}
is found.  Keeping terms in the symmetry energy up to quadratic order, one finds two conformal points
\begin{align}
    \alpha^{\text{conf.}}_{\pm}&=\dfrac{3}{5K_{\text{sym}}}\bigg[-\left(2L+K_{\text{sym}}\right)\label{eq:conformal_quadratic}\\
    &\pm\sqrt{\left(2L+K_{\text{sym}}\right)^2-10K_{\text{sym}}\left(L-J\right)}\bigg]\nonumber\\ &\text{[quadratic order in $S(n_B)$]}.\nonumber
\end{align}
The first key feature is that $n_B=n_0$ is a conformal point if $J=L$.  This is obvious from Eq.~\ref{eq:conformal_linear}, but still holds if the symmetry energy is extended to higher order (Eq.~\ref{eq:conformal_quadratic}).  There are many possible ways the symmetry energy could depend on density that give rise to no conformal points, as well.   
%%%%%%%%%%%%%%%%%%%%%%%%%%%%%%%%%%%%%%%%%%%%%%%%%%%%%%5
\section{Bulk viscosity}\label{sec:bulk_viscosity}
Bulk viscosity arises in systems in which an internal degree of freedom can be pushed out of equilibrium, but where partial reequilibration is able to occur.  In the dense nuclear matter in neutron stars, a density oscillation pushes a fluid element out of beta equilibrium, but the Urca process attempts to reestablish beta equilibrium.  

To calculate the bulk viscosity, consider a fluid element undergoing a harmonic density oscillation of (small) amplitude $\delta n_B$ and frequency $\omega$.  In response, the proton fraction will begin a harmonic oscillation too, but with a phase shift with respect to the density oscillation.  The size of the phase shift is a function of the oscillation amplitude and frequency, the susceptibilities of dense matter $A$ and $B$ (defined below), and the beta equilibration (Urca) rate $\gamma$.  The bulk viscosity is derived from this phase shift, which can be related to the work done on the fluid element as the density oscillation drives it to traverse a cycle in the pressure-volume plane \cite{Harris:2024evy}.  

Here, for simplicity, I focus on the subthermal bulk viscosity, where the oscillation is small enough that $\delta\mu\ll T$.  The restriction this imposes on $\delta n_B$ at a given temperature is shown in Fig.~6 of \cite{Harris:2024evy}.  The subthermal bulk viscosity is independent of the oscillation amplitude, which makes it a convenient regime to consider \cite{Alford:2010fd}.  The bulk viscosity in response to a density oscillation of frequency $\omega$ is given by \cite{Harris:2024evy}
 \begin{equation}
     \zeta\left(\omega\right) = \left\vert\dfrac{A^2}{B}\right\vert\dfrac{\gamma}{\omega^2+\gamma^2}. \label{eq:bv_formula}
 \end{equation}
The susceptibilities $A$ and $B$ are properties of the EoS, defined as
\begin{equation}
    A \equiv n_B\dfrac{\partial\delta\mu}{\partial n_B}\bigg\vert_{x_p},\quad
    B\equiv \dfrac{1}{n_B}\dfrac{\partial\delta\mu}{\partial x_p}\bigg\vert_{n_B}.\label{eq:A_and_B_definitions}
\end{equation}
The combination of susceptibilities in the prefactor of the bulk viscosity turns out to be related to the difference between the adiabatic and equilibrium incompressibilities of dense matter\footnote{$B$ is negative, so the adiabatic (fixed $x_p$) incompressibility is larger than the equilibrium (fixed $\delta\mu$) incompressibility.} \cite{Harris:2024ssp,Harris:2024evy,Lindblom:2001hd,Jones:2001ya,1987flme.book.....L}
\begin{align}
\dfrac{A^2}{B} &= -n_B  \dfrac{\partial P}{\partial x_p}\bigg\vert_{n_B}  \dfrac{\partial x_p}{\partial \delta\mu}\bigg\vert_{n_B}  \dfrac{\partial \delta\mu}{\partial n_B}\bigg\vert_{x_p}  \nonumber\\
&= n_B\dfrac{\partial P}{\partial n_B}\bigg\vert_{\delta\mu} - n_B\dfrac{\partial P}{\partial n_B}\bigg\vert_{x_p}.\label{eq:A2B_identity}
\end{align}

At fixed frequency $\omega$, the subthermal bulk viscosity is a function of density and temperature.  The susceptibility prefactor, and the susceptibilities themselves, depend on density but very little on temperature (c.f.~Fig.~3 in \cite{Alford:2019qtm}).  The beta-equilibration rate $\gamma$ depends strongly on temperature, due to the rapid opening up of phase space as the Fermi-Dirac distributions broaden with temperature, but depends relatively little on density\footnote{At low temperature, the direct Urca threshold in $npe$ matter introduces a sharp change in $\gamma$ at the density where $x_p=1/9$.  However, as temperature rises above about 1 MeV, the threshold blurs across density and even below the direct Urca threshold, direct Urca begins to dominate over modified Urca.  This effect lessens the density dependence \cite{Alford:2018lhf,Alford:2021ogv}.} (c.f.~Fig.~8 in \cite{Harris:2024evy} or Fig.~7 in \cite{Alford:2023uih}).

The $\gamma/\left(\omega^2+\gamma^2\right)$ factor in the bulk viscosity has a resonance structure and peaks as $\gamma$ approaches $\omega$.  Physically, this corresponds to actual value of the proton fraction $x_p$ being maximally out of phase with its beta-equilibrium value $x_p^{\text{eq.}}$, which leads to maximal $PdV$ work done and thus maximal bulk-viscous damping \cite{Harris:2024evy}.  To visualize the behavior of the bulk viscosity in the density-temperature plane, one relies on the fact that the susceptibility prefactor is mostly a function of density and the resonance factor is mostly a function of temperature.  At fixed density, the bulk viscosity is sharply peaked as a function of temperature, with the peak occurring at the temperature where $\gamma=\omega$.  The height of the peak is set, in part, by the susceptibility prefactor.  At the next density, the bulk viscosity as a function of temperature is peaked at roughly the same temperature, but the peak height may be different if $A^2/B$ has changed significantly.  Thus, the bulk viscosity has a ridge-like structure in density-temperature plan.  Contour plots of the bulk viscosity in the density-temperature plane are displayed in Fig.~4 in both \cite{Alford:2019qtm} and \cite{Alford:2023gxq}.  Also of interest is Fig.~2 in \cite{Most:2021zvc}, where the bulk viscosity is mapped on to a neutron star merger remnant.  

The susceptibilities are just properties of the EoS, and therefore they can be calculated within the framework of the symmetry energy expansion.  In fact, this has already been done \cite{Haensel:2000vz,Alford:2010gw,Alford:2010fd,Kolomeitsev:2014gfa,Tu:2025vew,Routray:2020zkf,Yang:2025yoo}, but without a focus on the details of the expansion, the particular values of the coefficients, or their possible correlations.  From Eqs.~\ref{eq:dmu_specifically} and \ref{eq:S(n)}, and then evaluating in beta equilibrium (Eq.~\ref{eq:beta_eq_condition}), the susceptibilities $A$ and $B$ are
\begin{widetext}
\begin{subequations}
\begin{align}
    A &= \dfrac{4}{3}\left(1-2x_p^{\text{eq.}}\right)\left[ L-J + \dfrac{1}{3}\left(2L+K_{\text{sym}}\right)\left(\dfrac{n_B-n_0}{n_0}\right)+\dfrac{1}{18}\left(5K_{\text{sym}}+Q_{\text{sym}}\right)\left(\dfrac{n_B-n_0}{n_0}\right)^2\right],\label{eq:Asusc}\\
    B &= -\dfrac{8}{n_0}\left[1+\dfrac{1}{6}\left(\dfrac{1-2x_p^{\text{eq.}}}{x_p^{\text{eq.}}}\right)\right]\left[J+\left(\dfrac{L}{3}-J\right)\left(\dfrac{n_B-n_0}{n_0}\right)+\left(J-\dfrac{L}{3}+\dfrac{K_{\text{sym}}}{18}\right)\left(\dfrac{n_B-n_0}{n_0}\right)^2\right].\label{eq:Bsusc}
\end{align}
\end{subequations}
Then the prefactor of the bulk viscosity, neglecting higher order terms in $x_p^{\text{eq.}}$, is
\begin{align}
    &\left\vert\dfrac{A^2}{B}\right\vert = \dfrac{4n_0}{3}x_p^{\text{eq.}}\left[1-8x_p^{\text{eq.}}+36(x_p^{\text{eq.}})^2\right]\bigg\{ \dfrac{\left(L-J\right)^2}{J}-\dfrac{L-J}{3J^2}\left(3J^2-2JK_{\text{sym}}-8JL+L^2\right)\left(\dfrac{n_B-n_0}{n_0}\right)\label{eq:A2B}\\
    &+\left[\dfrac{2 J^2 K_{\text{sym}}^2 - 30 J^3 L + 
  54 J^2 L^2 +JK_{\text{sym}}\left(36JL-5L^2-23J^2\right) - 18 J L^3 + 2 L^4 +2J^2\left(L-J\right)Q_{\text{sym}}}{18J^3}\right]\left(\dfrac{n_B-n_0}{n_0}\right)^2\bigg\}.\nonumber
\end{align}
\end{widetext}
The factor $\vert A^2/B\vert$ is exactly the factor $\zeta_0/\tau_{\Pi}$ calculated recently by Yang \textit{et al.}~\cite{Yang:2025yoo}.

Armed with an expression for $A^2/B$ as a function of density in terms of the symmetry energy coefficients $J$, $L$, $K_{\text{sym}}$, and $Q_{\text{sym}}$ ($x_p^{\text{eq.}}$ is just a function of these coefficients determined by solving Eq.~\ref{eq:beta_eq_condition}), I could now calculate the bulk viscosity $\zeta(\omega)$ with Eq.~\ref{eq:bv_formula}, if only I knew $\gamma(n_B,T)$.  Unfortunately, there is no way of calculating $\gamma$ purely in terms of the symmetry energy.  After all, it is a weak interaction, and the symmetry energy describes the dominant strong-interaction physics\footnote{The rate $\gamma$ is the susceptibility $B$ times the Urca rate.  The Urca rate, being a weak interaction, will naturally involve weak interaction parameters (e.g.~the Fermi coupling constant) that are negligible in the strong-interaction-dominated EoS.  Aspects of the strong interaction come into the Urca rate through the modified Urca process (and its recent improvements \cite{Alford:2024xfb}) and it would be interesting in the future to examine their relation to the symmetry energy.}.  However, extensive literature exists on the calculation of the Urca rate with Fermi's Golden Rule \cite{Lattimer:1991ib,Yakovlev:2000jp,Alford:2018lhf,Alford:2021ogv,Most:2022yhe,Tambe:2024usx,Kumamoto:2024jiq} or more advanced techniques \cite{Alford:2024xfb,Suleiman:2023bdf,Sedrakian:2004qd,Sedrakian:2007dj,Schmitt:2005wg}.  In the rest of the paper, I will focus on applications of the bulk viscosity where one does not need to know $\gamma$.  
%%%%%%%%%%%%%%%%%%%%%%%%%%%%%%%%%%%%%%%%%%%%%%%%%%%%%%%%%%%%%%%%%%%%%
\section{The peak value of bulk viscosity and its consequences}\label{sec:implications}
%%%%%%%%%%%%%%%%%%%%%%%%%%%%%%%%%%%%%%%%%%%%%%
\subsection{Peak value of bulk viscosity}
\begin{figure*}
  \centering
  \includegraphics[width=0.45\textwidth]{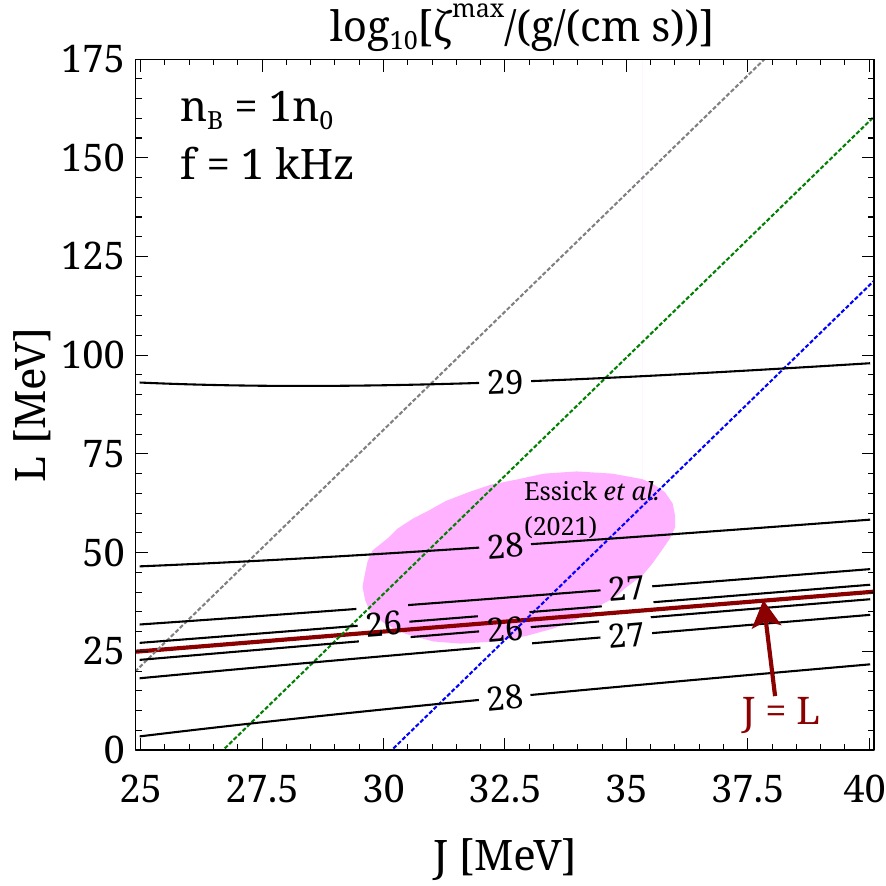}\quad\quad\quad
  \includegraphics[width=0.45\textwidth]{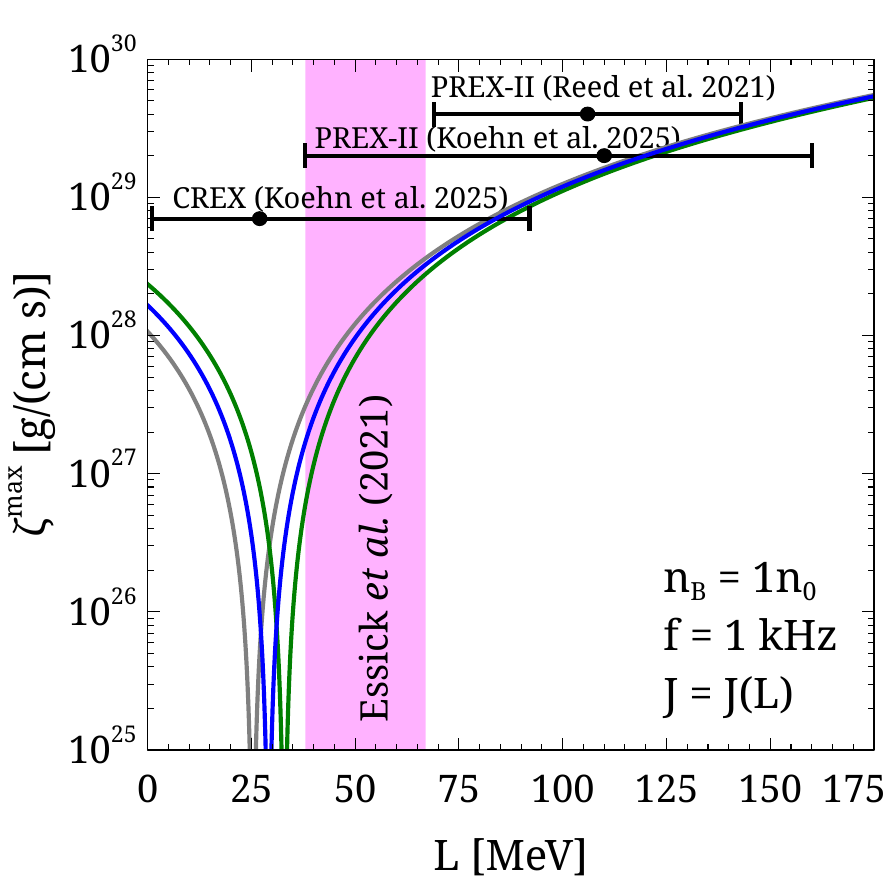}
  \caption{The maximum value of the bulk viscosity $\zeta^{\text{max}}$ at $1n_0$, for a 1 kHz density oscillation.  Left panel: $\zeta^{\text{max}}$ contour plot in the $JL$ plane.  Overlaid are the Essick \textit{et al.}~constraints (their Fig.~1, 90\% credible region) \cite{Essick:2021kjb} and the three $L = L(J)$ relations from Eq.~\ref{eq:L_of_J}.  Right panel: $\zeta^{\text{max}}$ as a function of $L$, using each of the three $L=L(J)$ lines, with the Essick \textit{et al.}, PREX-II,  and CREX constraints overlaid.  For PREX-II, I show the original Reed \textit{et al.}~constraint \cite{Reed:2021nqk} and the reanalysis by Koehn \textit{et al.}~\cite{Koehn:2024set}.  For CREX, I show just the analysis by Koehn \textit{et al.}~\cite{Koehn:2024set}.  The black dots indicate the median values.}
  \label{fig:bvmax_JL}
\end{figure*}
At a fixed density, the bulk viscosity forms a resonant peak structure.  The peak height, which I will call the maximum bulk viscosity - a function of density - is then
\begin{equation}
    \zeta^{\text{max}}\left(n_B\right) = \dfrac{1}{2\omega}\left\vert\dfrac{A^2}{B}\right\vert.\label{eq:zetamax}
\end{equation}
The value of the bulk viscosity at the peak is set by the EoS (and the hydrodynamics through $\omega$).  With Eq.~\ref{eq:A2B}, one can calculate the maximum bulk viscosity as a function of density.  The simplest expression is the maximum value of the bulk viscosity at $n_B=n_0$, which is
\begin{equation}
    \zeta^{\text{max}}\left(n_0\right) = \dfrac{2}{3}\dfrac{n_0}{\omega}x_p^{\text{eq.}}\left[1-8x_p^{\text{eq.}}+36\left(x_p^{\text{eq.}}\right)^2\right]\dfrac{\left(L-J\right)^2}{J}.\label{eq:bv_max_1n0}
\end{equation}
The proton fraction in beta equilibrium at $n_0$ is determined solely by the symmetry coefficient $J$, so $\zeta^{\text{max}}(n_0)$ depends only on $J$ and $L$ (for a given $\omega$).  

The first apparent feature is that the bulk viscosity vanishes at $n_0$ if $J=L$. 
 This is because, as discussed earlier, this is a conformal point. 
 If $J=L$, the proton fraction in beta equilibrium has a local maximum at $n_B=n_0$ and if one compresses a fluid element at $n_0$, to first order $x_p^{\text{eq.}}$ has not changed, and so the system is still in beta equilibrium, and thus there is no reequilibration needed and no bulk viscosity.

The left panel of Fig.~\ref{fig:bvmax_JL} displays a contour plot of the maximum bulk viscosity at $n_0$ in the $JL$ plane.  A density oscillation with a linear frequency $f=1\text{ kHz}$ (which is a typical frequency expected in neutron star mergers \cite{Alford:2017rxf}) is assumed. First, it is clear that the bulk viscosity sharply dips to zero along the line $J=L$.  As one departs from the line $J=L$, the value of $\zeta^{\text{max}}$ grows.  In particular, the maximum bulk viscosity is strongly dependent on the value of $L$, which is interesting given the controversy surrounding the value of $L$ at present.  Overlaid on the plot in pink is the 90\% credible region for $J$ and $L$ from Essick \textit{et al.}~\cite{Essick:2021kjb}.  Also overlaid are the three correlation lines (Eq.~\ref{eq:L_of_J}): green represents the median value and the gray and blue curves encompass 95.4\% of the EoSs as discussed near Eq.~\ref{eq:L_of_J}.  To avoid confusion, I note that the results of Essick \textit{et al.}~and the three correlation lines from Tews \textit{et al.}~are unrelated: the former is determined from a Bayesian analysis of $\chi$EFT, astrophysical measurements, and PREX-II data, while the latter represents correlations discovered within a large set of physically acceptable Skyrme and RMF EoSs.  

On the right hand side of Fig.~\ref{fig:bvmax_JL}, I plot the bulk viscosity as a function of $L$, with $J$ as a function of $L$ taking the values given in the 3 Tews \textit{et al.}~lines in the left panel.  Also displayed is the 90\% credible region for $L$ from Essick \textit{et al.}  The left panel indicates that the Essick \textit{et al.}~results allow for the existence of a conformal point at $n_B=n_0$.  However, once $J$ is marginalized over in the Essick \textit{et al.}~analysis, and once my $J$ values are ``collapsed'' onto $L$ values with the three correlations in Eq.~\ref{eq:L_of_J}, the right panel indicates that a conformal point at $n_B=n_0$ is unlikely.  Of course, more sophisticated analyses of the likelihood of a conformal point should be done.

The right panel of Fig.~\ref{fig:bvmax_JL} also shows the predicted values of the symmetry energy slope $L$ coming from PREX-II and CREX.  The PREX-II values, from Reed \textit{et al.}~and later from Koehn \textit{et al.}, predict a wide range for $L$, but the median values are quite high, around 100 MeV.  In contrast, CREX predicts a much lower value of $L$.  If PREX's median values are correct, than the bulk viscosity at saturation is large, around $10^{29}$ g/(cm s), while if the true value of $L$ is around the CREX median value, a conformal point of the EoS at $n_B=n_0$ is possible and the bulk viscosity could be quite small.  Thus, the tension between PREX and CREX has significant implications for the bulk viscosity at densities near saturation density.  However, if the more conventional values of $L$ are correct (represented by the Essick \textit{et al.}~credible interval), then the bulk viscosity at saturation density is likely around $10^{27}$ or $10^{28}$ g/(cm s).  Overall, the picture I present here is consistent with the results of Yang \textit{et al.}~\cite{Yang:2025yoo}, who found that small changes in $L$ result in large changes in the bulk viscosity.

\begin{figure}
  \centering
  \includegraphics[width=0.45\textwidth]{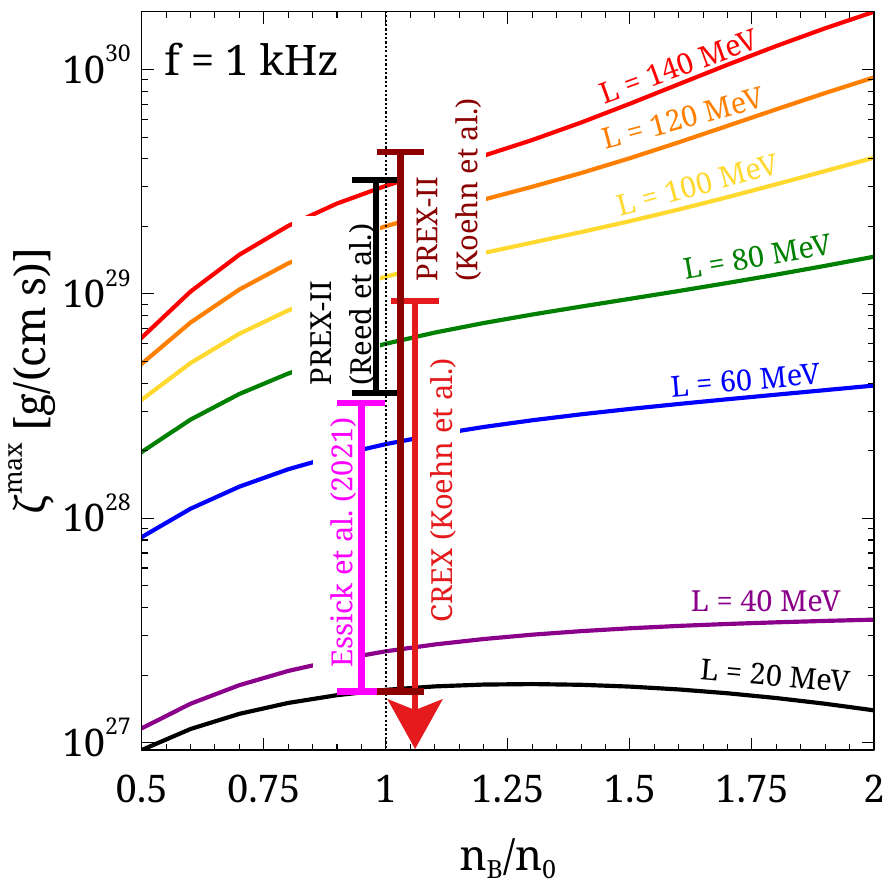}
  \caption{The resonant-maximum value of the bulk viscosity $\zeta^{\text{max}}$ as a function of density, assuming a 1 kHz density oscillation.  The different curves correspond to various choices of $L$.  The other symmetry coefficients $J$, $K_{\text{sym}}$, and $Q_{\text{sym}}$ are chosen according to the correlations Eqs.~\ref{eq:L_of_J}, \ref{eq:Ksym_of_L}, and \ref{eq:Qsym_of_L} using the median values.  Overlaid are the ranges of $\zeta^{\text{max}}$ predicted at $n_0$ from the values of $L$ found by Essick \textit{et al.}~\cite{Essick:2021kjb}, and the PREX-II \cite{Reed:2021nqk,Koehn:2024set} and CREX \cite{Koehn:2024set} data.}
  \label{fig:bvmax_vs_nB}
\end{figure}

It is also interesting to look at how the peak bulk viscosity changes as density increases.  Eqs.~\ref{eq:A2B} and \ref{eq:zetamax} indicate that $K_{\text{sym}}$ and $Q_{\text{sym}}$ appear in the $\zeta^{\text{max}}$ expression if one pushes away from $n_B=n_0$.  In Fig.~\ref{fig:bvmax_vs_nB}, I plot the maximum bulk viscosity as a function of density, for a few chosen values of $L$.  The other symmetry coefficients are chosen according to Eqs.~\ref{eq:L_of_J}, \ref{eq:Ksym_of_L}, and \ref{eq:Qsym_of_L}.  Also shown are the Essick \textit{et al.}, PREX-II, and CREX predictions for the bulk viscosity at $n_B=n_0$, given the range of $L$ values that they each predict.  

It is important to remember that the bulk viscosity as a function of density does not look like Fig.~\ref{fig:bvmax_vs_nB}.  This figure instead shows, if one plots the bulk viscosity in the density-temperature plane, the height of the peak if one traces it from low density to high density.  The actual behavior of $\zeta(n_B,T_0)$ at some fixed temperature $T_0$ depends on $\gamma$.  Fig.~\ref{fig:bvmax_vs_nB} shows that, generally, the bulk viscosity is larger as density increases.  The higher $L$ is, the faster this growth is.  Note that this is different from the observation made in, for example, \cite{Haensel:1992zz,Vidana:2012ex}, which found that the bulk viscosity increases with $L$.  These works concluded, correctly, that the beta-equilibration rate increases with the proton fraction, which increases with $L$, and this increases the bulk viscosity (for temperatures below the resonant peak temperature).  In this paper, I am focusing only on the resonant-peak value of the bulk viscosity, which depends on $L$ (among other symmetry parameters) due to the behavior of the susceptibilities and does not involve the beta-equilibration rate directly.  

In any case, as density increases the bulk viscosity likely (but not necessarily) exceeds the values shown in Fig.~\ref{fig:bvmax_JL}.  Calculations of the bulk viscosity for several RMF models in \cite{Alford:2019qtm,Alford:2023gxq,Harris:2024evy} confirm the tendency of the maximum bulk viscosity to rise with density.  Again, the value of $L$ is crucial in determining the behavior of the peak bulk viscosity value as a function of density.
%%%%%%%%%%%%%%%%%%%%%
\subsection{Damping timescale for local density oscillations}
\begin{figure}
  \centering
  \includegraphics[width=0.45\textwidth]{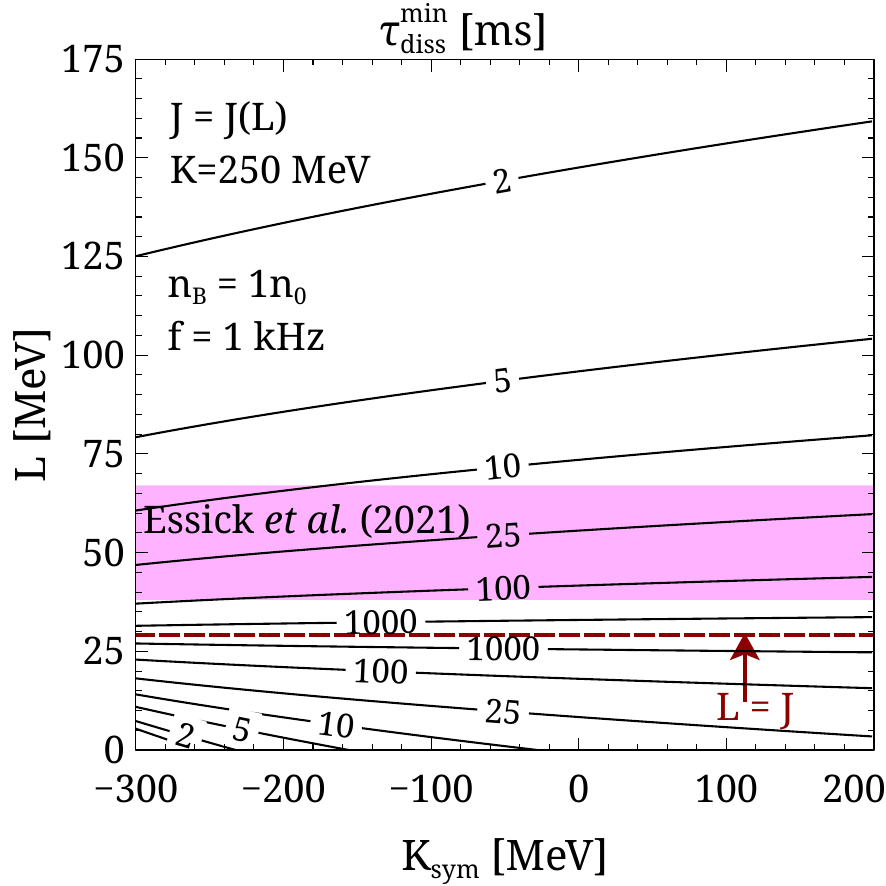}
  \caption{Minimum bulk-viscous damping time at $1n_0$, plotted in the $K_{\text{sym}}L$ plane.  Overlaid is the 90\% credible region for $L$ from the Essick \textit{et al.}~analysis \cite{Essick:2021kjb}.}
  \label{fig:tdissmin}
\end{figure}
A useful measure of the effect of bulk viscosity is the timescale for bulk viscosity to damp a small-amplitude density oscillation.  The amplitude of the oscillation cancels out, which makes this quantity particularly useful.  The energy density of an oscillation in nuclear matter is given by\footnote{One could presumably choose, for the purpose of estimating the
oscillation energy, to keep $\delta\mu$ fixed instead of $x_p$.  The difference between the adiabatic and equilibrium incompressibilities is small compared to either one of them (c.f.~Eq.\ref{eq:taudisscompressibility}).}
\begin{equation}
    \varepsilon = \dfrac{1}{2}\dfrac{\partial^2\varepsilon}{\partial n_B^2}\bigg\vert_{x_p}\left(\delta n_B\right)^2 = \dfrac{1}{2}\dfrac{\partial P}{\partial n_B}\bigg\vert_{x_p}\left(\dfrac{\delta n_B}{n_B}\right)\delta n_B
\end{equation}
and bulk viscosity dissipates energy density at a rate
\begin{equation}
    \dfrac{\mathop{d\varepsilon}}{\mathop{dt}}=\dfrac{\omega^2}{2}\left(\dfrac{\delta n_B}{n_B}\right)^2\zeta.
\end{equation}
Thus, a damping timescale can be obtained by dividing these two quantities \cite{Alford:2019qtm}
\begin{equation}
    \tau_{\text{diss}}=\dfrac{\varepsilon}{\mathop{d\varepsilon}/\mathop{dt}} = \dfrac{n_B\dfrac{\partial P}{\partial n_B}\bigg\vert_{x_p}}{\omega^2\zeta}.
\end{equation}
Naturally, at a given density, the fastest damping occurs when the bulk viscosity is maximum.  So, the minimum bulk-viscous damping time at a given density is also independent of the beta-equilibration rate $\gamma$.  It can be written
\begin{equation}
    \tau_{\text{diss}}^{\text{min}} = \dfrac{n_B\dfrac{\partial P}{\partial n_B}\bigg\vert_{x_p}}{\omega^2\zeta_{\text{max}}} = \dfrac{2}{\omega}\left(\dfrac{\dfrac{\partial P}{\partial n_B}\bigg\vert_{x_p}}{\dfrac{\partial P}{\partial n_B}\bigg\vert_{x_p} - \dfrac{\partial P}{\partial n_B}\bigg\vert_{\delta\mu}}\right),\label{eq:taudisscompressibility}
\end{equation}
where Eq.~\ref{eq:zetamax} and \ref{eq:A2B_identity} were used.
The incompressibility is the density derivative of the pressure (Eq.~\ref{eq:Pressure}) and is evaluated in beta equilibrium, yielding
\begin{widetext}
\begin{align}
\dfrac{\partial^2 \varepsilon}{\partial n_B^2}\bigg\vert_{x_p}&=\dfrac{1}{n_B}\dfrac{\partial P}{\partial n_B}\bigg\vert_{x_p} = \dfrac{1}{9n_0}\bigg\{ \left(K+K_{\text{sym}}+6L\right)\left(1+\dfrac{12J-24L-4K_{\text{sym}}}{K+K_{\text{sym}}+6L}x_p^{\text{eq.}}+...\right)  \\
&+ \dfrac{1}{3}\left(9K+9K_{\text{sym}}+Q+Q_{\text{sym}}\right)\left[ 1-4\left(\dfrac{9J-3L+9K_{\text{sym}}+Q_{\text{sym}}}{9K+9K_{\text{sym}}+Q+Q_{\text{sym}}}\right)x_p^{\text{eq.}}+...    \right] \left(\dfrac{n_B-n_0}{n_0}\right)       \bigg\}.\nonumber
\end{align}
\end{widetext}
This second derivative is essentially the spring constant of dense matter.  Notably, while for symmetric matter without leptons (e.g., nuclei), this spring constant would be just $K$, in beta-equilibrated dense matter, the relevant quantity becomes $K+K_{\text{sym}}+6L$, which can be significantly different from K (for high $L$, double or triple the size).

Therefore, oscillations in dense matter are damped from bulk viscosity on timescales as small as (suppressing higher orders in the proton fraction)
\begin{align}
    &\tau_{\text{diss}}^{\text{min}}\left(n_0\right)=\dfrac{1}{6\omega} \dfrac{J\left(K+K_{\text{sym}}+6L\right)}{(L-J)^2}\dfrac{1}{x_p^{\text{eq.}}}\\
    &\times \left\{1+4\left(\dfrac{3J+6L+K_{\text{sym}}+2K}{K+K_{\text{sym}}+6L}\right)x_p^{\text{eq.}}\right\}.\nonumber
\end{align}
For simplicity, I display only the value of $\tau_{\text{diss}}^{\text{min}}$ at $n_0$, though one can calculate it to linear order in $(n_B-n_0)/n_0$ without introducing higher order powers than $Q_{\text{sym}}$ in the symmetry energy.  

In Fig.~\ref{fig:tdissmin}, I plot the minimum damping time in the $K_{\text{sym}}L$ plane, with $J$ chosen according to Eq.~\ref{eq:L_of_J} and $K=250\text{ MeV}$.  The key feature here is that the damping time is the inverse of the bulk viscosity.  The bulk viscosity vanishes (at $n_0$) when $J = L$ and thus the damping time goes to infinity.  As $L$ increases, the maximum bulk viscosity grows dramatically and thus the damping time shrinks.  Superimposed on this is the effect that matter with higher $L$ and $K_{\text{sym}}$ is stiffer, and therefore an oscillation of a certain amplitude contains more energy than that same oscillation would in softer matter and thus the dissipation time is longer for the same bulk viscosity.  The overall message of this plot is that unless $L$ is quite close to $J$, rapid bulk-viscous damping can occur in neutron stars mergers if there are fluid elements that experience bulk viscosity at the resonant maximum (which itself is a key question for the relevance of bulk viscosity in neutron star mergers \cite{Most:2021zvc,Celora:2022nbp,Espino:2023dei}).  The Essick \textit{et al.}~constraints depicted in Fig.~\ref{fig:tdissmin} indicate that the damping times could be just under 10 ms, but could also be as long as several hundred ms.  Of course, if the PREX-II median values of $L$ are correct, then damping could be as quick as a few ms, whereas if the CREX median value is correct, then the damping time would be quite long.
%%%%%%%%%%%%%%%%%%%%%%%%%%%%%%%%%%%%%%%%%%%%%%%%%%%%%%%%%%%%%5
\subsection{Peak rotational frequency of neutron stars and the r-mode instability window}
Rotating neutron stars sustain oscillation modes that are not present in their nonrotating counterparts.  A famous example are the r-modes, with frequencies proportional to the rotation rate of the star \cite{Andersson:2024amk,Andersson:2019yve}.  In the absence of viscosity, the r-mode is unstable, and the rotation of the star is transferred into a growing r-mode amplitude, eventually spinning down the star.  However, viscosity can halt the growth of the r-mode.

The growth rate of the gravitational-wave instability rises rapidly with the rotation rate $\Omega$ of the star.  The damping timescales due to viscosity depend on the temperature of the matter in the star and the rotation rate.  Therefore, one can map out the r-mode instability landscape in the $T\Omega$ plane.  The boundary between stability and instability of the r-mode is found by solving
\begin{equation}
\tau_{\text{GW}}^{-1}\left(\Omega\right)+\tau_{\text{shear}}^{-1}\left(T\right)+\tau_{\text{bulk}}^{-1}\left(\Omega,T\right)=0\label{eq:GW_shear_bulk_equals0}
\end{equation}
for $\Omega(T)$.  Here, $\tau$ is the dissipation timescale (negative for gravitational waves, because they enhance the mode, and positive for the viscosities because they damp it).  

\begin{figure}
  \centering
  \includegraphics[width=0.45\textwidth]{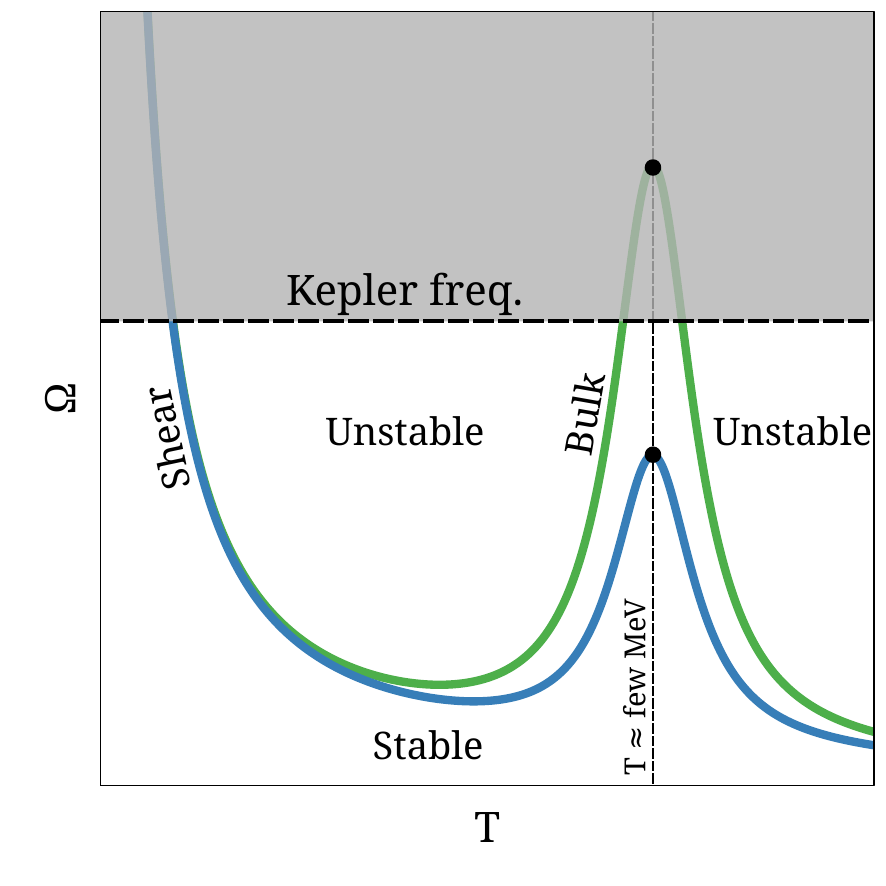}
  \caption{Schematic of the r-mode instability region.  The green and blue lines represent the boundary between stability and instability of the r-mode, where the destabilizing gravitational wave emission is opposed by shear viscosity at low temperature and bulk viscosity at high temperature.  Two possible strengths of the bulk viscosity are depicted in the green and blue curves.  Stars are unable to spin faster than the Kepler frequency.}
  \label{fig:r_mode_schematic}
\end{figure}

Fig.~\ref{fig:r_mode_schematic} shows a schematic of this $T\Omega$ plane.  Along the x-axis, $\Omega=0$ and the star in question is stable, as no r-modes are generated.  Even for very small $\Omega$, as long as there is any viscosity at all, the rotating star is stable.  However, as the rotation frequency grows larger, the gravitational-instability timescale drops dramatically and the star will be unstable.  Clearly, there is some border between stable and unstable stars, and two contrasting possibilities (for purely $npe$-matter stars) are sketched in blue and green in Fig.~\ref{fig:r_mode_schematic}.  The existence of hyperonic, quark, or superfluid phases, or dark matter, inside the neutron star leads to more complicated features \cite{2010MNRAS.408.1897H,Haskell:2012vg,Alford:2010fd,Alford:2019oge,Shirke:2023ktu}, but I do not consider these possibilities here.

As for the viscosities, the two contributions come from shear viscosity and from bulk viscosity.  The shear viscosity drops with temperature \cite{Andersson:2024amk}, while the bulk viscosity increases with temperature until it reaches a resonant peak at some fairly high temperature, and then decreases again.  Therefore, at low temperature, it is the shear viscosity that opposes the growth of the r-mode amplitude, but as temperature rises, the bulk viscosity supplants the shear viscosity as the stabilizing influence.  In this work, I will focus on the regime where the bulk viscosity dominates, and one can neglect the shear term in Eq.~\ref{eq:GW_shear_bulk_equals0}.  

A star cannot spin arbitrarily fast.  Above the Kepler frequency, the star begins to eject matter on its surface.  For a star with mass $M$ and radius $R$, the Kepler frequency is, to good approximation \cite{Glendenning:1997wn},
\begin{equation}
\Omega_K \approx 0.65 \sqrt{\dfrac{MG}{R^3}}.
\end{equation}
It only makes sense to consider stars rotating below the Kepler frequency, and thus the $\Omega>\Omega_K$ region in Fig.~\ref{fig:r_mode_schematic} is grayed out.  

The $l=m=2$ r-mode, which I focus on here because it dominates the gravitational wave emission, has a frequency of 
\begin{equation}
    \omega_r \approx \dfrac{2}{3}\Omega
\end{equation}
in the corotating frame.  In the inertial frame, the r-mode frequency is approximately $(4/3)\Omega$, but it is the corotating frame frequency $\omega_r$ that the fluid elements feel, and that is the frequency scale of relevance for the bulk viscosity in the rotating star.  

Since the Urca rate $\gamma$ depends relatively little on density, an isothermal neutron star at the right temperature could experience resonant bulk viscosity across its entire volume, with the value of the bulk viscosity being $\zeta(r) = \zeta^{\text{max}}(r)$ with $\zeta^{\text{max}}$ given by Eq.~\ref{eq:zetamax}.  That is, at this one temperature, the bulk viscosity in the star is entirely a function of $A^2/B$ and $\omega_r$.  A star at this temperature experiences more bulk-viscous dissipation than a star at any other temperature and thus this temperature supports the largest neutron star rotation rate $\Omega$ that is stable to r-modes.  This idea is depicted in Fig.~\ref{fig:r_mode_schematic}, where the two black dots correspond to the maximally-spinning neutron star discussed above, in the case where the maximum spin rate is less than (blue curve) or greater than (green curve) the Kepler frequency.  I have not found any definitive statement in the literature about which case is realized in nature, though most papers seem to assume the green curve scenario.  My goal here is to use the results from previous sections to determine whether the maximum stable rotation rate is above or below the Kepler frequency, as a function of the symmetry energy parameters.  

Since the focus here is on conditions in which the shear viscosity is subdominant, the boundary of stability is found by $\vert\tau_{\text{GW}}\vert=\vert\tau_{\text{bulk}}\vert$.  The gravitational radiation instability timescale for the $l=m=2$ mode is \cite{Alford:2010fd}
\begin{equation}
    \tau_{\text{GW}}^{-1} = -\dfrac{131072\pi}{164025}G\Omega^6\int_0^{R}\mathop{dr}r^6\varepsilon(r).\label{eq:tau_GW_rmode}
\end{equation}
Assuming the oscillations have small enough amplitude that the bulk viscosity remains subthermal, the timescale for bulk-viscous dissipation is (again, for $l=m=2$)
\begin{equation}
    \tau_{\text{bulk}}^{-1} = \dfrac{32}{1701}\dfrac{R^6\Omega^4}{\int_0^R\mathop{dr}r^6\varepsilon(r)}\int_0^R\mathop{dr}r^2\dfrac{1}{c_s^2}\zeta\left(\omega_r\right)\left[\delta\Sigma(r)\right]^2,
\end{equation}
where $c_s^2$ is the speed of sound and $\delta\Sigma(r)$ is an angle-averaged form of the density fluctuation of an r-mode (see \cite{Alford:2010fd} for full details and a lowest-order calculation of $\delta\Sigma$).  Essentially, the bulk-viscous dissipation time is an integral over the star of the bulk viscosity weighted by the r-mode profile.  The r-mode is stronger close to the edge of the star, probing densities of $1-2 n_0$.  As mentioned above, there will be a particular temperature where the entire star, if isothermal\footnote{In reality, an isothermal star consists of fluid elements at different local temperatures, due to redshift effects.  However, as the r-mode lies near the neutron star surface, I expect redshift effects to be relatively minor.}, is experiencing bulk viscosity at its resonant-peak value.  Thus, the bulk-viscous damping timescale at this one temperature becomes
\begin{equation}
    \tau_{\text{bulk}}^{-1} = \dfrac{8}{567}\dfrac{R^6\Omega^3}{\int_0^R\mathop{dr}r^6\varepsilon(r)}\int_0^R\mathop{dr}r^2\dfrac{1}{c_s^2}\bigg\vert\dfrac{A^2}{B}\bigg\vert\left[\delta\Sigma(r)\right]^2,\label{eq:tau_bulk_rmode}
\end{equation}
Therefore, the maximum of the instability boundary for r-modes can be found by setting $\vert\tau_{\text{GW}}\vert=\vert\tau_{\text{bulk}}\vert$ with the expressions Eq.~\ref{eq:tau_GW_rmode} and Eq.~\ref{eq:tau_bulk_rmode}, resulting in an equation for $\Omega_{\text{max}}^3$ in terms of, ultimately, the symmetry parameters.  It also depends on the mass of the star, of course, and I assume a $1.4M_{\odot}$ star for the following calculation.  Rather than use the symmetry energy expansion for $\varepsilon(r)$ and $c_s^2$, I just calculate them with a standard RMF EoS, but I have confirmed that the choice of EoS does not have much impact on the results, as the r-mode probes low densities where realistic EoSs mostly agree.  
\begin{figure}
  \centering
  \includegraphics[width=0.45\textwidth]{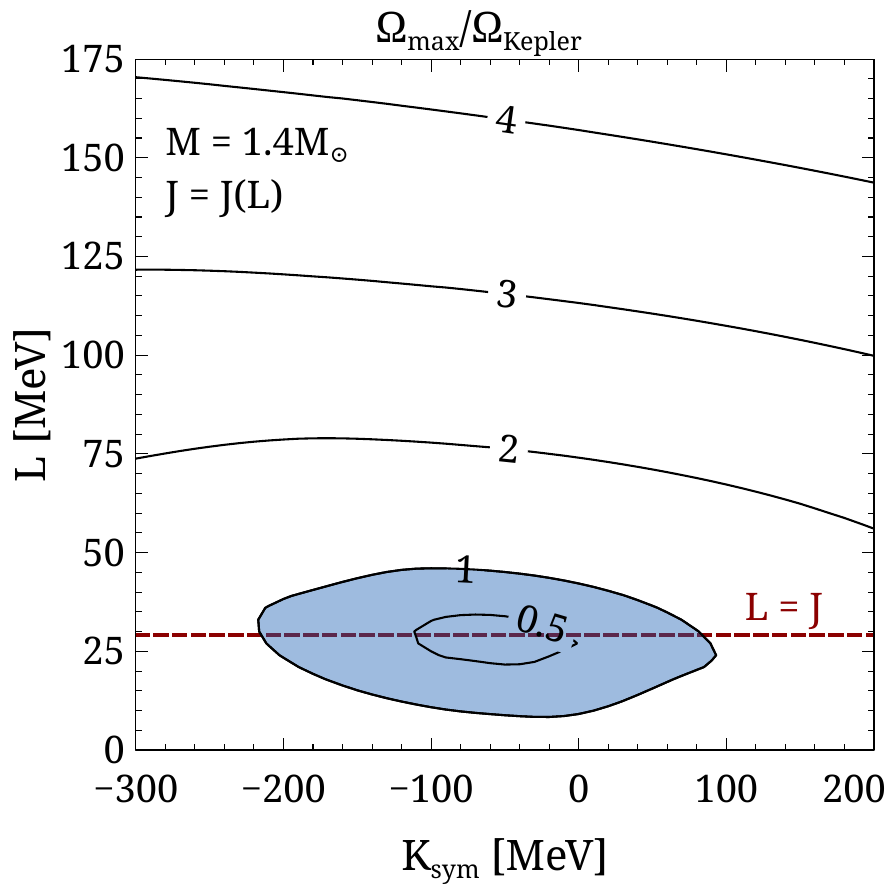}
  \caption{Contour plot of the maximum stable rotation frequency of a $1.4M_{\odot}$ $npe$-matter neutron star, normalized by the Kepler frequency, as a function of symmetry energy parameters $L$ and $K_{\text{sym}}$.  $J$ is set according to Eq.~\ref{eq:L_of_J}.  Only bulk-viscous damping is considered.  The shaded parameter space corresponds to an r-mode instability window such as that shown by the blue curve in Fig.~\ref{fig:r_mode_schematic}, while the unshaded parameter space has an instability window like that in the green curve in Fig.~\ref{fig:r_mode_schematic}.}
  \label{fig:JKL_omegapeak}
\end{figure}

In Fig.~\ref{fig:JKL_omegapeak}, I show a contour plot of the maximum spin frequency of a neutron star, normalized by the Kepler frequency, as a function of $L$ and $K_{\text{sym}}$.  The symmetry coefficient $J$ is chosen according to Eq.~\ref{eq:L_of_J}.  Small bulk viscosities correspond to small maximum stable rotation rates, and large bulk viscosities correspond to large maximum stable rotation rates, often above the Kepler frequency and thus not physically realizable.  Fig.~\ref{fig:JKL_omegapeak} indicates that there is a sizable pocket of symmetry energy parameter space that gives rise to a star with a maximum bulk viscosity that is still quite small, and thus only allows a star to spin stably at 100\%, 50\%, or even as low as 20\% of its Kepler frequency (the blue region in this plot corresponds to the blue curve in Fig.~\ref{fig:r_mode_schematic}).  This is consistent with the results of Figs.~\ref{fig:bvmax_JL} and \ref{fig:bvmax_vs_nB} which indicate that very small resonant-peak values for bulk viscosity are still allowed by experiment.  

The temperature at which this local maximum in the maximum stable rotation rate occurs cannot be determined by the EoS, just as the temperature where the bulk viscosity reaches its resonant maximum could not.  Calculations indicate in $npe$ matter, for a 1 kHz oscillation, the resonance occurs between temperatures of 2 and 6 MeV \cite{Alford:2019qtm,Alford:2023gxq}.  Depending on the mass of the star, and its rotation rate, the (corotating) r-mode frequency $\omega_r$ would not be 1 kHz, but instead only a few hundred Hz.  However, this only results in a slight ($\sim$ 1 MeV) decrease in the temperature of the resonance. 

At such high temperatures, while one can define an r-mode instability window within the paradigm discussed in this work (a rigidly rotating, isothermal neutron star made purely of $npe$ matter), complications come into the picture.  A piece of $npe$ matter, cooling via modified Urca neutrino emission, cools to below 1 MeV in just a few seconds (see the simple model developed in \cite{Page:2005fq}).  The timescale for the growth of the r-mode due to gravitational radiation is as small as 10 s (Eq.~\ref{eq:tau_GW_rmode} or see Fig.~4 of \cite{Alford:2010fd}).  A star located just above one of the two black dots in Fig.~\ref{fig:r_mode_schematic} will, in all likelihood, cool before it feels the instability.  In addition, neutron stars at temperatures this high are likely not isothermal, as the thermal equilibration of a neutron star core takes a decade or more \cite{Gnedin:2000me,Ho:2011aa}.  However, a non-isothermal star would just have a smaller fraction of its volume experiencing maximal bulk viscosity than the extreme case considered here.  Finally, at temperatures above several MeV, neutrinos become trapped and locally equilibrated with the $npe$ matter \cite{Alford:2018lhf,Alford:2019kdw} and thus the EoS must include their contribution as well.     
%%%%%%%%%%%%%%%%%%%%%%%%%%%%%%%%%%%%%%%%%%%%%%%%%%%%%
\section{Conclusions}
Parameterizing the EoS by expanding the symmetry energy around nuclear saturation density has become a standard technique in studying the EoS near saturation density.  I have computed the susceptibilities of $npe$ matter - key components of the subthermal bulk viscosity - in terms of the symmetry energy expansion coefficients (Eqs~\ref{eq:Asusc} and \ref{eq:Bsusc}).  I found that the susceptibility $A$, which describes the dependence of the pressure on the proton fraction, vanishes at saturation density if $J=L$, as this condition causes the proton fraction reaches a local maximum at saturation density.  Furthermore, I derived an equation (Eq.~\ref{eq:conformalpointseqn}) involving the symmetry energy and its derivative which can be solved to yield the densities where conformal points occur in the EoS.  

I used the expressions for the susceptibilities to calculate the resonant maximum value of the bulk viscosity at saturation density, as this quantity is independent of the beta-equilibration rate.  As expected, it vanishes if $J=L$, as it should at a conformal point.  I found that the resonant maximum value of the bulk viscosity increases strongly with $L$.  Current constraints on the symmetry parameters $J$ and $L$ indicate that the maximum bulk viscosity at $n_0$ could be quite small, but could also be as large as a few times $10^{28}$ g/(cm s).  If the PREX-II median value for $L$ is correct, then the peak value of the bulk viscosity would be a few times larger.  The larger $L$ is, the faster the peak value of bulk viscosity grows with increasing density.  The maximum bulk viscosity corresponds to a minimum bulk-viscous dissipation timescale, which I find at $n_0$ can be less than 10 ms within the allowed values of $L$.  

I apply the expression of the bulk viscosity in terms of the symmetry energy to the study of the r-mode instability window.  The maximum r-mode-stable stable spin rate of a neutron star as a function of temperature is expected to peak, and I find that for small values of $L$ and a relatively broad range of $K_{\text{sym}}$, this peak could lie below the Kepler frequency, meaning that a rapidly rotating neutron star containing only $npe$ matter could cool from extremely high temperatures and never pass through a temperature where the bulk viscosity would be high enough to stabilize the r-modes.  

The entirety of this paper applies only to neutron stars constructed of $npe$ matter.  A natural correction is to include muons.  Muons add another term to the energy per baryon (Eq.~\ref{eq:edens}), and their mass cannot be neglected.  However, they are still a free Fermi gas, and don't modify the symmetry energy expansion itself. 
 More fundamentally, they add another equilibration channel, where not only does the proton fraction equilibrate by the Urca process, but the muon fraction equilibrates by the muon-Urca process (where the electron is replaced by a muon) and also by the decay of muons to electrons.  This leads to, in principle, two peaks in the bulk viscosity instead of just one.  However the peaks are not, in practice, completely separate and their interaction makes it difficult to have a simple expression for the maximum of the bulk viscosity \cite{Alford:2021lpp,Alford:2022ufz,Alford:2023uih,Harris:2024evy}.  Recently, an effort was made to construct a symmetry energy expansion including strangeness \cite{Yang:2025wop,Bedaque:2014ada}.  It would be interesting to explore whether conclusions about the hyperonic bulk viscosity \cite{Alford:2020pld,Ofengeim:2019fjy} can be made. 
 Extensions of the work in this paper to suprathermal bulk viscosity would also be interesting, and of relevance to the saturation of the r-mode at large amplitudes \cite{Alford:2011pi}.  Finally, ultimately the r-mode is a general-relativistic phenomenon, and yet all of the calculations of the mode amplitude profile I relied on were Newtonian.  A relativistic study would be a natural upgrade \cite{Kraav:2024cus}.  
%%%%%%%%%%%%%%%%%%%%%%%%%%%%%%%%%%%
\section*{Acknowledgments}
I thank Chuck Horowitz for useful discussions and Jorge Noronha for showing me an advanced copy of his recent work \cite{Yang:2025yoo}.  I acknowledge the support of the National Science Foundation grant PHY 21-16686. 
\bibliography{references}
\end{document}